\documentclass[letter,journal]{IEEEtran}

\usepackage[dvipsnames]{xcolor}

\IEEEoverridecommandlockouts
\usepackage[utf8]{inputenc}
\usepackage[cmex10]{amsmath}

\usepackage{cleveref}
\usepackage{amsthm}
\usepackage{subfigure}
\usepackage{booktabs}
\usepackage{amsfonts}
\usepackage{amssymb}
\usepackage{bbm}
\usepackage{tikz}
\usepackage{multirow}

\usepackage{mathtools}
\usepackage[noadjust]{cite}
\usepackage{textcomp}
\usetikzlibrary{shapes,arrows}
\usepackage{comment}
\usepackage[noend]{algpseudocode}
\usepackage{algpseudocode}
\usepackage[linesnumbered,ruled,boxed]{algorithm2e}
\usepackage{caption}
\usepackage{tikz}

\usepackage{colortbl}

\usepackage{multicol}

\newtheorem*{theorem*}{Theorem}

\DeclareMathOperator{\tr}{tr}

\DeclareMathOperator{\re}{Re}

\title{\Large Electromagnetically Consistent Optimization Algorithms for the Global Design of RIS}
\author{M. W. Shabir, M. Di Renzo, \IEEEmembership{Fellow,~IEEE}, A. Zappone, \IEEEmembership{Senior~Member,~IEEE}, and M. Debbah, \IEEEmembership{Fellow,~IEEE} \vspace{-0.90cm}
\thanks{Manuscript received Sep. 25, 2024. M. W. Shabir, M. Di Renzo are with Universit\'e Paris-Saclay, CNRS, CentraleSup\'elec, Laboratoire des Signaux et Syst\`emes, Gif-sur-Yvette, France. (marco.di-renzo@universite-paris-saclay.fr). A. Zappone is with University of Cassino, Cassino, Italy. M. Debbah is with Khalifa University, Abu Dhabi, UAE. This work was supported by the European Commission under grant H2020 MetaWireless (956256). The work of M. Di Renzo was supported in part under grants HE COVER (101086228), HE UNITE (101129618), HE INSTINCT (101139161); France 2030 ANR-PEPR Networks of the Future (NF-YACARI 22-PEFT-0005); CHIST-ERA PASSIONATE (CHIST-ERA-22-WAI-04, ANR-23-CHR4-0003-01).}
\markboth{IEEE Wireless Communications Letters} {W. Shabir, et al., Electromagnetically Consistent Optimization Algorithms for the Global Design of RIS}
}
 
\begin{document}

\maketitle

\begin{abstract}
The reconfigurable intelligent surface is an emerging technology for wireless communications. We model it as an inhomogeneous boundary of surface impedance, and consider various optimization problems that offer different tradeoffs in terms of performance and implementation complexity. The considered non-convex optimization problems are reformulated as a sequence of approximating linear quadratically constrained or semidefinite programs, which are proved to have a polynomial complexity and to converge monotonically in the objective value. 
\end{abstract}
\vspace{-0.50cm}

\vspace{-0.75cm}
\section{Introduction}
\IEEEPARstart{T}{he} reconfigurable intelligent surface (RIS) is a physical layer technology that allows information and communication providers to optimize the propagation of electromagnetic waves, hence sculpting favorable communication channels for the efficient transmission and processing of information. Several optimization algorithms for RIS-aided systems are available, but most of them rely on simple (often simplistic) communication models \cite{MDR_PIEEE}. The development of electromagnetically consistent and tractable communication models for evaluating the performance and optimizing RIS-aided wireless networks, from a signal-level and system-level perspective, is, on the other hand, an important subject open to research \cite{ESITbits}.

In communication engineering, in addition, current optimization criteria for RIS-aided channels are usually based on the so-called local design, i.e., a specified constraint is imposed to each reconfigurable element of the RIS \cite{RobertGlobalDesignEE}. In the electromagnetic community, however, the local design criterion is known to be sub-optimal in terms of scattered power \cite{GlobalDesign}, and to possibly result in scattered electromagnetic waves towards directions different from the intended one \cite{MDR_PIEEE}, \cite{TAP_Floquet}, \cite{RadiationMaskVechi}. To overcome these limitations, an RIS needs to be optimized based on the so-called global design \cite{MDR_PIEEE}, \cite{GlobalDesign}, which ensures that the total power reflected towards the direction of interest is as close as possible (ideally equal) to the total incident power. This is ensured by imposing a single reflection constraint that encompasses all the reconfigurable elements of the RIS simultaneously. Existing optimization algorithms for the global design of an RIS are, however, based on communication models that are either not electromagnetically consistent \cite{RobertGlobalDesignEE} or rely on general-purpose optimization functions with no performance guarantee in terms of computational complexity, convergence properties, and optimality of the solution \cite{MDR_PIEEE}, \cite{LoadedArrays_FMINCOM}.

In this context, we embrace the electromagnetically consistent communication model introduced in \cite{MDR_PIEEE}, which models an RIS as an inhomogeneous boundary of surface impedance. Accordingly, we formulate several optimization problems with the following distinguishing features: (i) electromagnetic consistency of the solution; (ii) specified power efficiency towards the intended direction of reflection; (iii) specified maximum power towards unwanted directions of reflection; and (iv) specified physical implementation constraints, which are all imposed by design in the problem formulation. The considered optimization problems are shown not to be convex. To tackle them efficiently, we reformulate them as a sequence of approximating linear quadratically constrained or semidefinite programs, which are proved to have a polynomial complexity and to converge monotonically in the objective value \cite{OR_SequentialProgramming}. 

The significance of the proposed optimization algorithms is that the global design solution for anomalous reflectors is known only under ideal assumptions, i.e., unitary power efficiency, no parasitic scattering, and no physical constraints on the surface impedance, whose real part needs to be positive and negative \cite{GlobalDesign}. To the best of our knowledge, there exist no optimization algorithms to the design of RISs for which the power efficiency, undesired reradiations, and implementation constrains (e.g., the real part of the surface impedance shall not be negative) are specified as optimization constraints.

\textit{Notation}: Matrices and column vectors are denoted by bold uppercase and lowercase fonts. $|\cdot|$, ${(\cdot)}^*$, $\re(\cdot)$ denote the absolute value, conjugate, real part. ${(\cdot)}^{H}$, ${(\cdot)}^{T}$, $\tr(\cdot)$ denote the hermitian, transpose, trace operators. $\lVert \cdot \rVert$, $\lVert \cdot \rVert_*$, $\lVert \cdot \rVert_{\rm{F}}$ denote the spectral, nuclear, Frobenius norms. $\succeq$ denotes positive semidefinite. ${O}(\cdot)$ stands for the big-O notation. $j$ is the imaginary unit. ${\bf{1}}$ is the all ones vector. $\nabla \left( {f\left( {\bf{x}} \right)} \right)$ denotes the gradient of ${f\left( {\bf{x}} \right)}$ with respect to ${{{\bf{x}}^*}}$. $\nabla_n \left( {f\left( {\bf{x}} \right)} \right)$ is the $n$th entry of $\nabla \left( {f\left( {\bf{x}} \right)} \right)$. $f\left( {\bf{x }} \right) \approx f\left( {{\bar{\bf{x }}}} \right) + 2\re \left( \nabla^T\left( {f\left( \bar {\bf{x }} \right)} \right) \left( {{\bf{x }} - {\bar{\bf{x }}}} \right)^{*} \right)$ is the  first-order Taylor approximation of $f\left( {\bf{x }} \right)$ at the point $\bar{\bf{x}}$.

\vspace{-0.25cm}
\section{Electromagnetic Model}
\subsection{RIS Model}
We consider the same system model as in \cite[Sec. III-A]{MDR_PIEEE}, which encompasses a single-antenna transmitter, a single-antenna receiver, and an RIS (a flat surface $\mathcal{S}$) that is modeled as an inhomogeneous boundary of surface impedance with negligible thickness with respect to the considered wavelength. The RIS is modeled as a rectangle that lies in the $xy$-plane (i.e., $z = 0$) with its center located at the origin. Specifically, $\mathcal{S}$ is defined as $\mathcal{S} = \left\{(x,y) : |x|\leq L_x,|y|\leq L_y\right\}$, with $2L_x$ and $2L_y$ being the lengths of $\mathcal{S}$ along the $x$-axis and $y$-axis, respectively. We consider a reflecting RIS, i.e., the transmitter and receiver are located on the same side of $\mathcal{S}$.

The transmitter and receiver are located in the Fraunhofer far-field region of each other and of the RIS. Thus, the incident and reflected signals are modeled as plane waves, whose angles of incidence and reflection, with respect to the normal (i.e., the $z$-axis) to $\mathcal{S}$, are denoted by $\theta_i$ and $\theta_r$, respectively. As in \cite{MDR_PIEEE}, the incident and reflected signals propagate in the $yz$-plane, so that the dependence on the azimuth angle is ignored. We consider only the signal reflected by the RIS and ignore the transmitter-receiver direct link due to the presence of blocking objects. A free space propagation environment is considered.

Since the RIS is modeled as an inhomogeneous boundary of surface impedance, it is characterized by a surface impedance $Z(x,y)$ or, equivalently, by a surface reflection coefficient $\Gamma(x,y)$ for $(x,y) \in \mathcal{S}$, according to the definitions in \cite[Eq. (37)]{MDR_PIEEE}. Because of the independence from the azimuth angle, $Z(x,y)$ and $\Gamma(x,y)$ are constant functions along the $x$-axis. For system optimization, we can hence consider $Z(x,y) = Z(y)$ and $\Gamma(x,y) = \Gamma(y)$. For ease of writing and numerical implementation when solving the optimization problems, the surface impedance and reflection coefficient are discretized with spatial sampling $\Delta_x$ and $\Delta_y$ along the $x$-axis and $y$-axis, respectively, as detailed in \cite[Sec. III-C]{MDR_PIEEE}. Accordingly, $Z(y)$ and $\Gamma(y)$ are represented by two column vectors $\mathbf{z} = [z_{1},z_{2},\dots,z_{N}]^T$ and $\boldsymbol{\gamma} = [\gamma_{1},\gamma_{2},\dots,\gamma_{N}]^T$, respectively. Specifically, $z_{n} = Z(y_n)$ with  $y_n = -L_y - \Delta_y/2 + n\Delta_y$, $n=1,2,\ldots,N$, and $N=2L_y/\Delta_y$. The same applies to $\boldsymbol{\gamma}$.

The entries of $\mathbf{z}$ and $\boldsymbol{\gamma}$ are related to one another as follows: \vspace{-0.1cm}
\begin{equation}
  \label{Surface_impedance_definition}
    z_n = \eta_0 \frac{1 + {\gamma_n}}{\cos \theta_i - {\gamma_n} \cos \theta_r}, \quad \gamma_n=\frac{{z_n}\cos\theta_{i}-\eta_{0}}{{z_n}\cos\theta_{r}+\eta_{0}} \vspace{-0.1cm}
\end{equation}
where $\eta_0$ is the free space impedance and $n=1,2,\ldots,N$.

\vspace{-0.25cm}
\subsection{Electromagnetic Consistency}
The entries of $\mathbf{z}$ and $\boldsymbol{\gamma}$ are arbitrary complex values, with the only requirement that they need to produce reflected electric and magnetic fields, given the incident electric and magnetic fields, that are electromagnetically consistent, i.e., that fulfill Maxwell's equations \cite[p. 1183]{MDR_PIEEE}. This implies that the entries of $\boldsymbol{\gamma}$ need to satisfy the constraint \cite[Eq. (44), Table 5]{MDR_PIEEE} \vspace{-0.1cm}
\begin{equation}
H_{n}\left(\boldsymbol{\gamma}\right)=\frac{\left|f_{n}^{\prime\prime}-2j \kappa f_{n}^{\prime}\sin\theta_{r}\right|}{\kappa^{2}\left|\gamma_{n}\right|} = 0, \quad n=1, \ldots, N-2 \label{hz_basic definition} \vspace{-0.1cm}
\end{equation}
where $f_{n}=\gamma_{n}e^{j\kappa\left(\sin\theta_{r}-\sin\theta_{i}\right)y_{n}}$, $f_{n}^{\prime}=({f_{n+1}-f_{n}})/{\Delta_{y}}$, $f_{n}^{\prime\prime}=({f_{n+1}^{\prime}-f_{n}^{\prime}})/{\Delta_{y}}$, $\kappa=2\pi/\lambda$, and $\lambda$ is the considered wavelength. The corresponding constraint that the entries of $\mathbf{z}$ need to fulfill can be found by inserting $\gamma_n$ in \eqref{Surface_impedance_definition} into \eqref{hz_basic definition}. 

As detailed in \cite{MDR_PIEEE}, the condition $H_{n} = 0$ results in the optimal solution, at the highest implementation complexity, as the surface impedance is characterized by large variations, and by positive and negative values of its real part, which make it difficult to implement the resulting RIS in practice \cite{GlobalDesign}, \cite{LoadedArrays_FMINCOM}. For this reason, the constraint $H_{n} = 0$ that ensures the electromagnetic consistency of the solution is usually relaxed, by replacing it with the constraint $\varepsilon_{\rm L} \le H_{n} \le \varepsilon_{\rm U}$, where $0 \le \varepsilon_{\rm L} \le \varepsilon_{\rm U}$ are small positive constants that control the tradeoff between the optimality and electromagnetically consistency (i.e., $\varepsilon_{\rm U} \to 0$) of the obtained design against the implementation complexity (i.e., $\varepsilon_{\rm L} > 0$) of the RIS.

\vspace{-0.25cm}
\subsection{Performance Metrics}
As detailed in \cite[Sec. III-B]{MDR_PIEEE}, the performance of an RIS is completely characterized by two performance metrics.

\textbf{Surface Net Power Flow} -- The surface net power flow is defined as the difference between the power reradiated by the whole RIS towards the intended direction of reflection and the total incident power. If the power reradiated towards the intended direction of reflection is hence equal to the total incident power, the surface net power flow is zero. An RIS for which the surface net power flow is equal to zero is defined as globally optimum. To optimize an RIS according to the global design criterion, the surface net power flow needs then to be zero. Considering typical performance versus implementation tradeoffs, the global design criterion is hence tantamount to either minimizing the surface net power flow or ensuring that it is as close as possible to zero within a specified tolerance.

Based on the considered electromagnetic model, the surface net power flow can be formulated, in terms of $\boldsymbol{\gamma}$, as follows: \vspace{-0.1cm}
\begin{align}
\label{Surface_power_def}
P_{\mathcal{S}}(\boldsymbol{\gamma}) = {a_X}{\Delta _y}\left( {{c_i} + {\alpha _r}{{\boldsymbol{\gamma }}^H}{\boldsymbol{\gamma }} + 0.5{\alpha _{ir}}\left( {{{\bf{1}}^T}{\boldsymbol{\gamma }} + {{\boldsymbol{\gamma }}^H}{\bf{1}}} \right)} \right) \vspace{-0.1cm}
\end{align}
where $a_X={\left|E_{0}\right|^{2}L_{x}}/{\eta_{0}} \ge 0$, $\alpha_{i}=\cos\theta_{i} \ge 0$, $\alpha_{r}=\cos\theta_{r} \ge 0$, $c_{i}=-2L_{y}\alpha_{i}/{\Delta _y} \le 0$, $\alpha_{ir}=\alpha_{r}-\alpha_{i} \in [-1,+1]$, $E_{0}$ is the amplitude of the incident electric field (a plane wave).

\textbf{Power Flux} -- The power flux characterizes the amount of power reradiated by an RIS at a specified point of observation. We focus on observation points located in the far-field region of the RIS. The power flux provides information on the angular response of the RIS, i.e., how the total incident power is reradiated towards different directions. In the far-field, the power flux is proportional to the radiation pattern of the RIS. The power flux is an essential performance indicator in wireless communications since it determines the amount of received power and hence the signal-to-interference ratio.

Based on the considered electromagnetic model, the power flux evaluated towards a generic direction of reradiation $\theta_k$ can be formulated, as a function of $\boldsymbol{\gamma}$, as follows: \vspace{-0.1cm}
\begin{equation}
\label{Power_flux_def}
{P_{{\theta _k}}}\left( {\boldsymbol{\gamma }} \right) = {a_k}\Delta _y^2{\chi _{ik}}{\left| {{{\boldsymbol{\gamma }}^T}{\bf{u}}_{ik}} \right|^2} \vspace{-0.1cm}
\end{equation}
where the $n$th entry of vector ${\bf{u}}_{ik}$ is $u_{ik,n}=e^{j \kappa \left(\sin\theta_{k}-\sin\theta_{i}\right)y_{n}}$ for $n=1,2, \ldots, N$, ${a}_{k}=\frac{\kappa^{2}}{\eta_{0}}\frac{\left|E_{0}\right|^{2}L_{x}^{2}}{8\pi^{2}R_{k}^{2}}$, $R_k$ is the distance from the RIS towards the direction $\theta_k$, and ${\chi}_{ik}=\cos^{2}\theta_{r}+\cos^{2}{\theta}_{k}+2\cos\theta_{r}\cos{\theta}_{k} \ge 0$. If the direction of observation coincides with the intended direction of reflection, i.e., ${\theta}_{k} = {\theta}_{r}$, then $R_k = R_r$, ${\chi}_{ik} = {\chi}_{ir}= 4\cos^{2}\theta_{r}$, and $P_{{\theta}_{k}}\left(\boldsymbol{\gamma}\right) = P_{{\theta}_{r}}\left(\boldsymbol{\gamma}\right)$.

Next, we formulate and solve optimization problems that aim to minimize $P_{\mathcal{S}}(\boldsymbol{\gamma})$ in \eqref{Surface_power_def} (Sec. III) and to maximize $P_{{\theta}_{r}}\left(\boldsymbol{\gamma}\right)$ in \eqref{Power_flux_def} (Sec. IV) subject to specified design constraints.

\vspace{-0.25cm}
\section{Optimization: Surface Net Power Flow}
In \cite{GlobalDesign}, it is shown that minimizing $P_{\mathcal{S}}(\boldsymbol{\gamma})$ in \eqref{Surface_power_def} results in an optimal anomalous reflector that steers the total incident power towards the intended direction of reflection with no undesired scattering towards other directions. The optimal surface impedance $\bf{z}$ is known in a closed-form expression only if the constraint in \eqref{hz_basic definition} is strictly fulfilled, leading to a high implementation complexity \cite{GlobalDesign}, \cite{LoadedArrays_FMINCOM}. To the best of our knowledge, there exists no general and efficient optimization framework that aims to minimize $P_{\mathcal{S}}(\boldsymbol{\gamma})$ in \eqref{Surface_power_def} by imposing specified design constraints. This is tackled in this section.

\vspace{-0.25cm}
\subsection{Global Design -- Helmholtz Constraint}
We commence generalizing \cite{GlobalDesign}, by relaxing the constraint in \eqref{hz_basic definition} within the specified tolerances $\varepsilon_{\rm{HC_L}} \ge 0$ and $\varepsilon_{\rm{HC_U}} \ge 0$. This keeps under control the variations of the surface impedance $\bf{z}$ along $\mathcal{S}$, facilitating the implementation of the resulting RIS, while ensuring a quasi-Maxwellian solution \cite{MDR_PIEEE}.

The considered problem can be stated as follows: \vspace{-0.15cm}
\begin{align}
(\textbf{S-HC}) \quad  & \underset{\boldsymbol{\gamma}}{\min} \quad \left| P_{\mathcal{S}}(\boldsymbol{\gamma}) \right|  \label{Eq:S-HC} \\
\text{s.t.} \quad  &\mathcal{H}_{n}(\boldsymbol{\gamma}) \leq \varepsilon_{\rm{HC_U}}, \quad n=1, 2, \ldots, N-2 \hspace{1.3cm} \text{(a)} \nonumber \\ 
 &\mathcal{H}_{n}(\boldsymbol{\gamma}) \geq \varepsilon_{\rm{HC_L}}, \quad n=1, 2, \ldots, N-2 \hspace{1.3cm} \text{(b)} \nonumber \vspace{-0.15cm}
\end{align}

The problem \textbf{S-HC} is not convex. To tackle it efficiently, the objective function in \eqref{Eq:S-HC} is rewritten in epigraph form \cite[Eq. (4.11)]{boyd2004convex}. Also, the function $\mathcal{H}_{n}(\boldsymbol{\gamma})$ in \eqref{Eq:S-HC} is formulated explicitly in terms of the optimization variable $\boldsymbol{\gamma}$, as follows: \vspace{-0.15cm}
\begin{equation}
\label{Eq:HC_Simplified}
\mathcal{H}_{n}(\boldsymbol{\gamma}) = \frac{{\left| {g_n}\left( {{\gamma _n},{\gamma _{n + 1}},{\gamma _{n + 2}}} \right)\right|}}{{{\kappa ^2}\Delta _y^2\left| {{\gamma _n}} \right|}} = \frac{{\left| {{g_n}\left( {\boldsymbol{\gamma }}_n \right)} \right|}}{{{\kappa ^2}\Delta _y^2\left| {{\gamma _n}} \right|}} \vspace{-0.15cm}
\end{equation}
where $\beta_{1}=1+2j \kappa \sin\theta_{r}\Delta_{y}$, $\beta_{2}=-\left(2+2j \kappa \sin\theta_{r}\Delta_{y}\right)$, and ${{g_n}\left( {\boldsymbol{\gamma }}_n \right)}  = {g_n}\left( {{\gamma _n},{\gamma _{n + 1}},{\gamma _{n + 2}}} \right) = {u_{n + 2}}{\gamma _{n + 2}} + {\beta _2}{u_{n + 1}}{\gamma _{n + 1}} + {\beta _1}{u_n}{\gamma _n}$ with ${{\boldsymbol{\gamma }}_n} = {\left( {{\gamma _n},{\gamma _{n + 1}},{\gamma _{n + 2}}} \right)^T}$.

The problem \textbf{S-HC} can then be rewritten as follows: \vspace{-0.15cm}
\begin{align}
& (\textbf{S-HC-a}) \quad 
\underset{\mathbf{\boldsymbol{\gamma}},t}{\min} \quad t \quad \text{s.t.} \quad -t \le 0 \; \text{(a)} \label{Eq:S-HC-a} \\ 
& P_{\mathcal{S}}(\boldsymbol{\gamma}) - t \le 0 \; \text{(b)},  \quad \left|{{g_n}\left( {\boldsymbol{\gamma }}_n \right)} \right| - {{\tilde \varepsilon }_{{\mathop{\rm {HC_U}}\nolimits} }}\left| {{\gamma _n}} \right| \le 0 \; \text{(c)} \nonumber  \\
& - P_{\mathcal{S}}(\boldsymbol{\gamma}) - t \le 0 \;  \text{(d)}, \quad -\left|{{g_n}\left( {\boldsymbol{\gamma }}_n \right)} \right| + {{\tilde \varepsilon }_{{\mathop{\rm {HC_L}}\nolimits} }}\left| {{\gamma _n}} \right| \le 0 \; \text{(e)} \nonumber \vspace{-0.15cm}
\end{align}
where ${{\tilde \varepsilon }_{{\mathop{\rm {HC_{L,U}}}\nolimits} }} = {\varepsilon _{{\mathop{\rm {HC_{L,U}}}\nolimits} }}{\kappa ^2}\Delta _y^2$ and $n=1, 2, \ldots, N-2$.

Problem \textbf{S-HC-a} is still not convex due to the constraints (\ref{Eq:S-HC-a}c)-(\ref{Eq:S-HC-a}e). To tackle it, we embrace the iterative inner approximation framework \cite{OR_SequentialProgramming}. Specifically, we consider convex upper bounds for the concave functions in (\ref{Eq:S-HC-a}c)-(\ref{Eq:S-HC-a}e), replacing them with their first-order Taylor approximation. As for the function ${{{\boldsymbol{\gamma }}^H}{\boldsymbol{\gamma }}}$, we note that $\nabla \left( {{{\boldsymbol{\gamma }}^H}{\boldsymbol{\gamma }}} \right) = {\boldsymbol{\gamma }}$ and introduce the following lower bound ${a_X}{\Delta _y}{\alpha _r}{{{{\boldsymbol \gamma }}}^H}{{\boldsymbol \gamma }} \ge {p_{\mathcal{S}}}\left( {\boldsymbol{\gamma }}, \bar {\boldsymbol{\gamma }} \right)$ at the point ${\bar{\boldsymbol \gamma }}$: \vspace{-0.15cm}
\begin{equation}
{p_{\mathcal{S}}}\left( {\boldsymbol{\gamma }}, \bar {\boldsymbol{\gamma }} \right) = {a_X}{\Delta _y}{\alpha _r}\left({{{\bar{\boldsymbol \gamma }}}^H}{\bar{\boldsymbol \gamma }} + 2 \re \left( {\nabla ^T}\left( {{\bar{\boldsymbol{\gamma }}^H}{\bar{\boldsymbol{\gamma }}}} \right)\left( {{\boldsymbol{\gamma }} - {\bar{\boldsymbol \gamma }}} \right)^{*} \right) \right) \nonumber \vspace{-0.15cm}
\end{equation}

As for the function $\left|{{g_n}\left( {\boldsymbol{\gamma }}_n \right)} \right|$, we note that ${\nabla _n}\left( {\left| {{g_n}\left( {{{\boldsymbol{\gamma }}_n}} \right)} \right|} \right) = {{0.5\beta _1^*u_n^*{g_n}\left( {{{\boldsymbol{\gamma }}_n}} \right)} \mathord{\left/
 {\vphantom {{0.5\beta _1^*u_n^*{g_n}\left( {{{\boldsymbol{\gamma }}_n}} \right)} {\left| {{g_n}\left( {{{\boldsymbol{\gamma }}_n}} \right)} \right|}}} \right. \kern-\nulldelimiterspace} {\left| {{g_n}\left( {{{\boldsymbol{\gamma }}_n}} \right)} \right|}}$ and introduce the following lower bound $\left| {{g_n}\left( {{{\boldsymbol{\gamma }}_n}} \right)} \right| \ge \left| {g_n^{\rm{L}}\left( {{{\boldsymbol{\gamma }}_n}} \right)} \right|$ evaluated at the point ${{{{\boldsymbol{\bar \gamma }}}_n}}$: \vspace{-0.15cm}
\begin{equation}
\left| {g_n^{\rm{L}}\left( {{{\boldsymbol{\gamma }}_n}} \right)} \right| = \left| {{g_n}\left( {{{{\bar{\boldsymbol \gamma }}}_n}} \right)} \right| + 2{\mathop{\rm Re}\nolimits} \left( {{\nabla _n}\left( {\left| {{g_n}\left( {{{{\bar{\boldsymbol \gamma }}}_n}} \right)} \right|} \right){{\left( {{\gamma _n} - {{\bar \gamma }_n}} \right)}^*}} \right) \nonumber \vspace{-0.15cm}
\end{equation}

Therefore, the following reformulation for \textbf{S-HC-a} is obtained at the generic iteration of the inner algorithm: \vspace{-0.15cm}
\begin{align}
& (\textbf{S-HC-b}) \quad  \underset{\mathbf{\boldsymbol{\gamma}},t}{\min} \quad t \quad \text{s.t.} \quad (\ref{Eq:S-HC-a}\text{a}), \; (\ref{Eq:S-HC-a}\text{b}) \label{Eq:S-HC-b} \\ 
& - {a_X}{\Delta _y}\left( {{c_i} + 0.5{\alpha _{ir}}\left( {{{\bf{1}}^T}{\boldsymbol{\gamma }} + {{\boldsymbol{\gamma }}^H}{\bf{1}}} \right)} \right) - {p_{\mathcal{S}}}\left( {\boldsymbol{\gamma }}, \bar {\boldsymbol{\gamma }} \right)  - t \le 0 \hspace{0.13cm} \text{(a)} \nonumber \\
&
\hspace{-0.2cm}\left|{g_n}\left( {{{{\boldsymbol{\gamma }}}_n}} \right)\right| - {{\tilde \varepsilon }_{{\mathop{\rm {HC_U}}\nolimits} }}\left( \left| {{{\bar \gamma }_n}} \right| +  2 \re \left(\nabla_n \left( {\left| {{{\bar \gamma }_n}} \right|} \right)\left( {{\gamma _n} - {{\bar \gamma }_n}} \right)^{*} \right) \right) \le 0 \; \text{(b)} \nonumber \\
& -\left|{{g_n^{\rm L}}\left( {\boldsymbol{\gamma }}_n \right)} \right| + {{\tilde \varepsilon }_{{\mathop{\rm {HC_L}}\nolimits} }}\left| {{\gamma _n}} \right| \le 0 \; \text{(c)}, \quad \left\| {{\boldsymbol{\gamma }} - {\bar{\boldsymbol \gamma }}} \right\| \le {\varepsilon _{{\rm{TR}}}} \; \text{(d)} \nonumber \vspace{-0.15cm}
\end{align}
where ${\bar{\boldsymbol{\gamma }}}$ is the point at which Taylor's approximation is made (the solution of the preceding iteration of the inner algorithm), ${\bar{\boldsymbol{\gamma }}_n} = {\left( {{{\bar \gamma }_n},{{\bar \gamma }_{n + 1}}, {{\bar \gamma }_{n + 2}}} \right)^T}$, and ${\nabla _n}\left( {\left| {{\gamma _n}} \right|} \right) = 0.5{{{\gamma _n}} \mathord{\left/ {\vphantom {{{\gamma _n}} {\left| {{\gamma _n}} \right|}}} \right. \kern-\nulldelimiterspace} {\left| {{\gamma _n}} \right|}}$. Eq. (\ref{Eq:S-HC-b}d) is the trust region, which ensures that, at each iteration, the set of feasible solutions is limited to the points for which the approximation is sufficiently accurate \cite{TrustRegion}. The radius of the trust region is controlled via the small positive constant ${\varepsilon _{{\rm{TR}}}}$, which is updated at each iteration as detailed in Sec. V.

\textbf{S-HC-b} is convex and is solved as detailed in Sec. III-D.

\vspace{-0.25cm}
\subsection{Global Design -- Reradiation Mask}
Fulfilling the constraint in \eqref{hz_basic definition} ensures the electromagnetic consistency of the obtained solution. The optimization problem has, however, a number of constraints that is equal to the number of optimization variables $N$. Motivated by recent results \cite{MDR_PIEEE}, \cite{TAP_Floquet}, \cite{RadiationMaskVechi}, \cite{LoadedArrays_FMINCOM}, we show that the constraint in \eqref{hz_basic definition} can be replaced by keeping the power scattered towards specified (undesired) directions of reradiation below a given maximum level. The advantage of this formulation is that the unwanted directions of reradiation may be known apriori \cite{TAP_Floquet},  \cite{TWC_StructuralScattering}, and their number may be much less than the number of optimization variables $N$. The resulting optimization constraint is referred to as reradiation mask constraint \cite{MDR_PIEEE}, \cite{RadiationMaskVechi}.

The considered problem can be stated as follows: \vspace{-0.1cm}
\begin{align}
(\textbf{S-RM}) \quad  &\underset{\boldsymbol{\gamma}}{\min} \quad \left| P_{\mathcal{S}}(\boldsymbol{\gamma}) \right|  \label{Eq:S-RM} \\
\text{s.t.} \quad &{P_{{\theta _k}}}\left( {\boldsymbol{\gamma }} \right) \le {\varepsilon _{{\rm{RM}}}}, \quad {\theta _k} \in \mathcal{K} \nonumber \vspace{-0.1cm}
\end{align}
where $\mathcal{K}$ is the set of undesired directions of radiation and ${\varepsilon _{{\rm{RM}}}}$ is the maximum amount of radiated power towards them.

The problem \textbf{S-RM} can be tackled efficiently by rewriting the objective function in epigraph form \cite[Eq. (4.11)]{boyd2004convex}, and by making explicit the constraint in (\ref{Eq:S-RM}) by using \eqref{Power_flux_def}, as follows: \vspace{-0.1cm}
\begin{align}
(\textbf{S-RM-a}) \quad  
& \underset{\mathbf{\boldsymbol{\gamma}},t}{\min} \quad t \quad \text{s.t.} \quad (\ref{Eq:S-HC-a}\text{a}), \; (\ref{Eq:S-HC-a}\text{b}), \; (\ref{Eq:S-HC-b}\text{a}), \; (\ref{Eq:S-HC-b}\text{d}) \label{Eq:S-RM-a} \\ 
& \left| {{\boldsymbol{\gamma }}^T}{{\bf{u}}_{ik}} \right| - \sqrt {{{\tilde \varepsilon }_{{\rm{RM}}}}} \le 0, \quad {\theta _k} \in \mathcal{K}  \hspace{1.70cm} (\text{a})  \nonumber \vspace{-0.1cm}
\end{align}
where ${{\tilde \varepsilon }_{{\rm{RM}}}} = \frac{{{\varepsilon _{{\rm{RM}}}}}}{{{a_k}\Delta _y^2{\chi _{ik}}}}$. \textbf{S-RM-a} is convex and can be solved efficiently as detailed in Sec. III-D.

\vspace{-0.25cm}
\subsection{Approximated Global Design -- Reactive Impedance}
The optimization problems \textbf{S-HC} and \textbf{S-RM} aim to maximize the radiation efficiency of the RIS, by simultaneously maximizing and minimizing the power reradiated towards the intended and undesired directions of reradiation, respectively. \textbf{S-HC} and \textbf{S-RM} impose, however, mild or no implementation constraint to the feasible set of solutions, respectively. As illustrated in \cite{MDR_PIEEE}, \cite{GlobalDesign}, the surface impedance $\bf{z}$ obtained by solving \textbf{P-HC} and \textbf{P-RM} has a negative real part, which makes the obtained design difficult to be implemented \cite{GlobalDesign}. Thus, we consider an optimization problem that imposes specified implementation constraints by design. A sought after implementation requirement is that the real part of $\bf{z}$ is not negative and as small as possible, ideally equal to zero, in order to minimize the power losses. Since the real part of $\bf{z}$ cannot take negative values, the obtained design cannot be deemed globally optimum but only approximately globally optimum.

The considered problem can be stated as follows: \vspace{-0.1cm}
\begin{align}
(\textbf{S-RI}) \quad  &\underset{\boldsymbol{\gamma}}{\min} \quad \left| P_{\mathcal{S}}(\boldsymbol{\gamma}) \right|  \label{Eq:S-RI} \\
\text{s.t.} \quad &{P_{{\theta _k}}}\left( {\boldsymbol{\gamma }} \right) \le {\varepsilon _{{\rm{RM}}}}, \quad {\theta _k} \in \mathcal{K} \hspace{2.95cm} \text{(a)} \nonumber \\
& \re \left( {{z}}_n \right) \ge 0, \quad \; \; \; n=1, 2, \ldots, N \hspace{1.8cm} \text{(b)} \nonumber  \\
& \re \left( {{z}}_n \right) \le  {\varepsilon _{{\rm{RI}}}}, \quad n=1, 2, \ldots, N \hspace{1.78cm} \text{(c)} \nonumber \vspace{-0.1cm}
\end{align} 
where ${\varepsilon _{{\rm{RI}}}} \ge 0$ is used to adjust the tradeoff between the power losses and  implementation complexity. If ${\varepsilon _{{\rm{RI}}}} =0$, the losses are zero but the implementation complexity is the highest.

The objective function in \eqref{Eq:S-RI} and the constraint in (\ref{Eq:S-RI}\text{a}) can be tackled as in \textbf{S-RM-a}. The constraints in (\ref{Eq:S-RI}\text{b}) and (\ref{Eq:S-RI}\text{c}) can be reformulated in terms of ${\boldsymbol{\gamma }}$ from \eqref{Surface_impedance_definition}. For each pair $(z_n,\gamma_n)$, (\ref{Eq:S-RI}\text{b}) and (\ref{Eq:S-RI}\text{c}) are equivalent to the following: \vspace{-0.5cm}
\begin{align} 
& \text{(\ref{Eq:S-RI}\text{b})}:\quad {\alpha _r}{\left| {{\gamma _n}} \right|^2} - \left( {{\alpha _i} - {\alpha _r}} \right)\re \left( {{\gamma _n}} \right) - {\alpha _i} \le 0 \label{Eq:ReZ_1} \\
& \text{(\ref{Eq:S-RI}\text{c})}:\quad - \left( {1 + {{\tilde \varepsilon }_{{\rm{RI}}}}{\alpha _r}} \right)  {\alpha _r} {\left| {{\gamma _n}} \right|^2} + {\alpha _i}\left( {1 - {{\tilde \varepsilon }_{{\rm{RI}}}}{\alpha _i}} \right) \nonumber  \vspace{-0.1cm}
\end{align}
\begin{align} 
& \hspace{1.5cm} + \left( {{\alpha _i} - {\alpha _r} + 2{{\tilde \varepsilon }_{{\rm{RI}}}}{\alpha _i}{\alpha _r}} \right){\mathop{\rm Re}\nolimits} \left( {{\gamma _n}} \right) \le 0 \label{Eq:ReZ_2}  \vspace{-0.1cm}
\end{align}
where ${{\tilde \varepsilon }_{{\rm{RI}}}} = {{{\varepsilon _{{\rm{RI}}}}} \mathord{\left/ {\vphantom {{{\varepsilon _{{\rm{RI}}}}} {{\eta _0}}}} \right. \kern-\nulldelimiterspace} {{\eta _0}}}$. The constraint in \eqref{Eq:ReZ_1} is convex. The constraint in \eqref{Eq:ReZ_2} is concave, and it is tackled through an upper bound obtained from Taylor's approximation applied to ${\left| {{\gamma _n}} \right|^2}$.

By introducing the function $\psi \left( {{\gamma _n}} \right) = {\alpha _i}\left( {1 - {{\tilde \varepsilon }_{{\rm{RI}}}}{\alpha _i}} \right) + \left( {{\alpha _i} - {\alpha _r} + 2{{\tilde \varepsilon }_{{\rm{RI}}}}{\alpha _i}{\alpha _r}} \right){\mathop{\rm Re}\nolimits} \left( {{\gamma _n}} \right)$ and considering $\nabla_n \left( {{{\boldsymbol{\gamma }}^H}{\boldsymbol{\gamma }}} \right) = {{\gamma}_n}$, \textbf{S-RI} can be reformulated as follows (for $n=1,2,\ldots,N$): \vspace{-0.5cm}
\begin{align}
& (\textbf{S-RI-a}) \quad  
\underset{\mathbf{\boldsymbol{\gamma}},t}{\min} \quad t \quad \text{s.t.} \; (\ref{Eq:S-HC-a}\text{a}), \; (\ref{Eq:S-HC-a}\text{b}), \; (\ref{Eq:S-HC-b}\text{a}), \; (\ref{Eq:S-HC-b}\text{d}), \; (\ref{Eq:S-RM-a}\text{a}), \; \eqref{Eq:ReZ_1} \nonumber \\ 
& \psi \left( {{\gamma _n}} \right) - {\tilde \alpha _r} \left( {\left| {{{\bar \gamma }_n}} \right|^2} + 2 \re \left( \nabla_n \left( {{\bar{\boldsymbol{\gamma }}^H}{\bar{\boldsymbol{\gamma }}}} \right)   \left( {{\gamma _n} - {{\bar \gamma }_n}} \right)^{*} \right) \right) \le 0 \nonumber \vspace{-0.1cm}
\end{align}
where ${\tilde \alpha _r}= {\left( {1 + {{\tilde \varepsilon }_{{\rm{RI}}}}{\alpha _r}} \right)  {\alpha _r}}$. The problem \textbf{S-RI-a} is convex and can be solved efficiently as detailed in Sec. III-D.

\vspace{-0.25cm}
\subsection{Convergence and Complexity}
The complete algorithm to solve \textbf{S-HC-b}, \textbf{S-RM-a}, and \textbf{S-RI-a} is an instance of sequential programming. The trust region is, in fact, applied in the neighborhood of $\bar{\boldsymbol{\gamma}}$, which is, by design, a feasible point at any iteration. Thus, the proposed algorithm generates a monotonically decreasing sequence of objective values that converges in the objective \cite{OR_SequentialProgramming}. The convex problem solved at each iteration has a linear objective and quadratic constraints, which can be tackled using the interior-point method. Thus, the total arithmetic cost per iteration is ${\mathcal{O}}\left( {c^{1/2}}\left( {c{v^2} + {v^3}} \right) \right)$ \cite[p. 232]{IPMs_Complexity}, with $v$ and $c$ being the numbers of optimization variables and constraints, respectively.

\vspace{-0.25cm}
\section{Optimization: Power Flux}
In electromagnetic theory, the surface net power flow is the optimality criterion usually used to design perfect anomalous reflectors \cite{GlobalDesign}. In  communication theory, the optimization problem is usually formulated by maximizing the power scattered towards the intended direction of reflection (known as the power flux), while imposing specified constraints to the power efficiency of the RIS \cite{RobertGlobalDesignEE}. The two problem formulations are naturally related to one another, but the corresponding solutions and algorithms are different. Often, the problem formulation considered in communication theory is more general, as (i) it can be extended to different objective functions, besides the reflected power, and scenarios, and (ii) it can account for the surface net power flow as an optimization constraint, resulting in specified performance versus implementation complexity tradeoffs by design \cite{RobertGlobalDesignEE}. However, the computational complexity and memory requirements of the power flux optimization problem are usually higher.

In this section, therefore, we analyze the globally optimum design of an RIS under the lenses of communication theory, by considering alternative formulations for \textbf{S-RM} and \textbf{S-RI}. \textbf{S-HC} is not considered for brevity and because it is \textit{de facto} equivalent to \textbf{S-RM} from a communication perspective.

\vspace{-0.25cm}
\subsection{Global Design -- Reradiation Mask}
By considering the power scattered towards the intended direction of reflection as the objective function and the surface net power flow as an optimization constraint, a problem \textit{de facto} equivalent to \textbf{S-RM} can be formulated as follows: \vspace{-0.1cm}
\begin{align}
& (\textbf{P-RM}) \quad  \underset{\boldsymbol{\gamma}}{\max} \quad {P_{{\theta _r}}}\left( {\boldsymbol{\gamma }} \right) 
\label{Eq:P-RM} \\
& \text{s.t.} \quad {P_{{\theta _k}}}\left( {\boldsymbol{\gamma }} \right) \le {\varepsilon _{{\rm{RM}}}}, \; {\theta _k} \in \mathcal{K} \; \text{(a)}, \quad  \left| P_{\mathcal{S}}(\boldsymbol{\gamma}) \right| \le \varepsilon_{\rm{SP}} \; \text{(b)}  \nonumber \vspace{-0.1cm}
\end{align}
where $\varepsilon_{\rm{SP}} \ge 0$ is used to adjust the tradeoff between the amount of power scattered towards the intended direction of reflection and the implementation complexity of the RIS.

\textbf{P-RM} is not convex. To tackle it, we introduce ${\boldsymbol{\Gamma }} = {\boldsymbol{\gamma }}{{\boldsymbol{\gamma }}^H}$, ${\bf{U}}_{ik} = {\bf{u}}_{ik}{{\bf{u}}_{ik}^H}$. Since ${{{\boldsymbol{\gamma }}^H}{\boldsymbol{\gamma }}} = \tr\left( {\boldsymbol{\Gamma }} \right)$ in \eqref{Surface_power_def}, we define \vspace{-0.1cm}
\begin{equation}
{{\hat P}_{\mathcal{S}}}\left( {{\boldsymbol{\Gamma }},{\boldsymbol{\gamma }}} \right) = {a_X}{\Delta _y}\left( {{c_i} + {\alpha _r}\tr\left( {\boldsymbol{\Gamma }} \right) + 0.5{\alpha _{ir}}\left( {{{\bf{1}}^T}{\boldsymbol{\gamma }} + {{\boldsymbol{\gamma }}^H}{\bf{1}}} \right)} \right) \nonumber \vspace{-0.1cm}
\end{equation}

Then, \textbf{P-RM} can be equivalently formulated as follows: \vspace{-0.1cm}
\setlength{\belowdisplayskip}{3pt}
\begin{align}
& (\textbf{P-RM-a}) \quad \underset{\boldsymbol{\Gamma}, \boldsymbol{\gamma}}{\max} \quad  {a_r}\Delta _y^2{\chi _{ir}} \tr \left( {{\boldsymbol{\Gamma}} {\bf{U}}_{ir}^*} \right)
\label{Eq:P-RM-a} \\
& \text{s.t.} \quad {a_k}\Delta _y^2{\chi _{ik}} \tr \left( {{\boldsymbol{\Gamma}} {\bf{U}}_{ik}^*} \right) - {\varepsilon _{{\rm{RM}}}}  \le 0, \quad {\theta _k} \in \mathcal{K} \hspace{0.45cm} \text{(a)} \nonumber \\
& \left|{\hat P}_{\mathcal{S}}(\boldsymbol{\Gamma}, \boldsymbol{\gamma}) \right| - \varepsilon_{\rm{SP}}  \le 0 \quad \text{(b)},  \quad {\boldsymbol{\Gamma }} = {\boldsymbol{\gamma }}{{\boldsymbol{\gamma }}^H} \quad \text{(c)} \nonumber \vspace{-0.15cm}
\end{align} 

The only non-convex constraint in \textbf{P-RM-a} is (\ref{Eq:P-RM-a}\text{c}). To tackle it, we reformulate it equivalently as follows: \vspace{-0.1cm}
\begin{align} 
& \text{(\ref{Eq:P-RM-a}\text{c})}:\; {\left\| {\boldsymbol{\Gamma }} \right\|_*} - {\left\| {\boldsymbol{\Gamma }} \right\|_{\rm{F}}} \le 0  \; \text{(a)} 
 \label{Eq:SDP_Rank1Constraint} \\
& \text{(\ref{Eq:P-RM-a}\text{c})}:\; \left[ {\begin{array}{*{20}{c}}
{\boldsymbol{\Gamma }}&{\boldsymbol{\gamma }}\\
{{{\boldsymbol{\gamma }}^H}}&1
\end{array}} \right] \succeq	 0 \; \text{(b)}, \quad {\left\| {\boldsymbol{\Gamma }} \right\|} - {\left\| {\boldsymbol{\gamma }} \right\|^2} \le 0 \; \text{(c)}  \nonumber \vspace{-0.1cm}
\end{align} 
where (\ref{Eq:SDP_Rank1Constraint}\text{a}) ensures that ${\boldsymbol{\Gamma }}$ has rank one and (\ref{Eq:SDP_Rank1Constraint}\text{b}) ensures that ${\boldsymbol{\Gamma }}$ is positive semidefinite. Also, (\ref{Eq:SDP_Rank1Constraint}\text{b}) and (\ref{Eq:SDP_Rank1Constraint}\text{c}) ensure that ${\left\| {\boldsymbol{\Gamma }} \right\|} - {\left\| {\boldsymbol{\gamma }} \right\|^2} =0$, i.e., the only singular value of ${\left\| {\boldsymbol{\Gamma }} \right\|}$ is equal to ${\left\| {\boldsymbol{\gamma }} \right\|^2}$. This is because the spectral norm is monotone, i.e., ${\boldsymbol{\Gamma }} \succeq {\boldsymbol{\gamma }}{{\boldsymbol{\gamma }}^H}$ in (\ref{Eq:SDP_Rank1Constraint}\text{b}) implies $\left\| {\boldsymbol{\Gamma }} \right\| \ge \left\| {{\boldsymbol{\gamma }}{{\boldsymbol{\gamma }}^H}} \right\| = {\left\| {\boldsymbol{\gamma }} \right\|^2}$.

The constraints in (\ref{Eq:SDP_Rank1Constraint}\text{a}) and (\ref{Eq:SDP_Rank1Constraint}\text{d}) are not convex, since they are given by the difference of two convex functions. We tackle them by applying the iterative inner approximation framework \cite{OR_SequentialProgramming}. Specifically, we consider the following convex lower bounds for the functions ${\left\| {\boldsymbol{\gamma }} \right\|^2}$ and ${\left\| {\boldsymbol{\Gamma }} \right\|_{\rm{F}}}$: \vspace{-0.1cm}
\begin{align} 
& {f_{\rm{V}}}\left( {\boldsymbol{\gamma }}, \bar { \boldsymbol{\gamma }} \right) = {\left\| {{\bar{\boldsymbol \gamma }}} \right\|^2} + 2{\mathop{\rm Re}\nolimits} \left( {{\nabla ^T}\left( {{{\left\| \bar {\boldsymbol{\gamma }} \right\|}^2}} \right)\left( {{\boldsymbol{\gamma }} - {\bar{\boldsymbol \gamma }}} \right)^{*}} \right) \label{Eq:TaylorNorm} \\
& {f_{\rm{F}}}\left( {\boldsymbol{\Gamma }}, \bar { \boldsymbol{\Gamma }} \right) = {\left\| {{\bar{\boldsymbol \Gamma }}} \right\|_{\rm{F}}} + 2{\mathop{\rm Re}\nolimits} \left( {\sum\nolimits_{n,m} \hspace{-0.15cm}{{\delta _{n,m}}\left( \bar {\boldsymbol{\Gamma }} \right){{\left( {{\Gamma _{n,m}} - {{\bar \Gamma }_{n,m}}} \right)}^*}} } \right) \nonumber \vspace{-0.1cm}
\end{align} 
with ${\left\| {\boldsymbol{\gamma }} \right\|^2} \ge {f_{\rm{V}}}\left( {\boldsymbol{\gamma }}, \bar { \boldsymbol{\gamma }} \right)$, ${\left\| {\boldsymbol{\Gamma }} \right\|_{\rm{F}}} \ge {f_{\rm{F}}}\left( {\boldsymbol{\Gamma }}, \bar { \boldsymbol{\Gamma }} \right)$, ${{\Gamma _{n,m}}}$ is the $(n,m)$th entry of ${\boldsymbol{\Gamma }}$, $\nabla \left( {{{\left\| {\boldsymbol{\gamma }} \right\|}^2}} \right) = {\boldsymbol{\gamma }}$, ${\delta _{n,m}}\left( {\boldsymbol{\Gamma }} \right) = 0.5{{{\Gamma _{n,m}}} \mathord{\left/ {\vphantom {{{\Gamma _{n,m}}} {{{\left\| {\boldsymbol{\Gamma }} \right\|}_{\rm{F}}}}}} \right. \kern-\nulldelimiterspace} {{{\left\| {\boldsymbol{\Gamma }} \right\|}_{\rm{F}}}}}$, and ${\bar{\boldsymbol{\Gamma }}}$ and ${\bar{\boldsymbol{\gamma }}}$ are the points at which Taylor's approximation is made (the solution of the preceding iteration).

At the generic iteration of the inner algorithm, therefore, \textbf{P-RM-a} is tackled by solving the following problem: \vspace{-0.1cm}
\begin{align}
& (\textbf{P-RM-b}) \quad \underset{\boldsymbol{\Gamma}, \boldsymbol{\gamma}}{\max} \quad  {a_r}\Delta _y^2{\chi _{ir}} \tr \left( {{\boldsymbol{\Gamma}} {\bf{U}}_{ir}^*} \right) \label{Eq:P-RM-b} \\
& \text{s.t.} \quad (\ref{Eq:P-RM-a}\text{a}), \; (\ref{Eq:P-RM-a}\text{b}), (\ref{Eq:SDP_Rank1Constraint}\text{b})  \quad \text{(a)} \nonumber \\
& {\left\| {\boldsymbol{\Gamma }} \right\|_{*}} - {f_{\rm{F}}}\left( {{\boldsymbol{\Gamma }},{\bar{\boldsymbol \Gamma }}} \right) \le 0 \quad \text{(b)}, \quad {\left\| {\boldsymbol{\Gamma }} \right\|} - {f_{\rm{V}}}\left( {{\boldsymbol{\gamma }},{\bar{\boldsymbol \gamma }}} \right) \le 0  \quad \text{(c)} \nonumber \\
& \left\| {{\boldsymbol{\gamma }} - {\bar{\boldsymbol \gamma }}} \right\| \le {\varepsilon _{{\rm{TR,}}{\boldsymbol{\gamma }}}} \quad \text{(d)}, \quad \left\| {{\boldsymbol{\Gamma }} - {\bar{\boldsymbol \Gamma }}} \right\| \le {\varepsilon _{{\rm{TR,}}{\boldsymbol{\Gamma }}}} \quad \text{(e)} \nonumber 
\vspace{-0.1cm}
\end{align}
where ${\varepsilon _{{\rm{TR,}}{\boldsymbol{\gamma }}}} \ge 0$, ${\varepsilon _{{\rm{TR,}}{\boldsymbol{\Gamma }}}} \ge 0$, and (\ref{Eq:P-RM-b}e) and (\ref{Eq:P-RM-b}d) are trust region constraints, similar to the constraint in (\ref{Eq:S-HC-b}d).

\textbf{P-RM-b} is convex and is solved as detailed in Sec. IV-C.

\vspace{-0.25cm}
\subsection{Approximated Global Design -- Reactive Impedance}
By considering the power scattered towards the intended direction of reflection as the objective function and the surface net power flow as an optimization constraint, a problem \textit{de facto} equivalent to \textbf{S-RI} can be formulated by adding the constraints in (\ref{Eq:S-RI}\text{b}) and (\ref{Eq:S-RI}\text{c}), or, equivalently, the constraints in \eqref{Eq:ReZ_1} and \eqref{Eq:ReZ_2}, to the problem  \textbf{P-RM-b} in \eqref{Eq:P-RM-b}. 

By setting ${\left| {{\gamma _n}} \right|^2} = {\Gamma _{n,n}}$, we then obtain ($n=1,2, \ldots,N$): \vspace{-0.6cm}
\begin{align}
(\textbf{P-RI}) \quad & \underset{\boldsymbol{\Gamma}, \boldsymbol{\gamma}}{\max} \quad  {a_r}\Delta _y^2{\chi _{ir}} \tr \left( {{\boldsymbol{\Gamma}} {\bf{U}}_{ir}^*} \right) \label{Eq:P-RI} \\
& \text{s.t.} \quad (\ref{Eq:P-RM-a}\text{a}), \; (\ref{Eq:P-RM-a}\text{b}), \;  (\ref{Eq:SDP_Rank1Constraint}\text{b}), \; (\ref{Eq:P-RM-b}\text{b})\text{-}(\ref{Eq:P-RM-b}\text{e})  \nonumber \\
& {\alpha _r}{\Gamma _{n,n}} - \left( {{\alpha _i} - {\alpha _r}} \right){\mathop{\rm Re}\nolimits} \left( {{\gamma _n}} \right) - {\alpha _i} \le 0 \nonumber \\
&  - \left( {1 + {{\tilde \varepsilon }_{{\rm{RI}}}}} \right){\alpha _r}{\Gamma _{n,n}} + {\alpha _i}\left( {1 - {{\tilde \varepsilon }_{{\rm{RI}}}}{\alpha _i}} \right)  \nonumber \\
& \hspace{1cm}+ \left( {{\alpha _i} - {\alpha _r} + 2{{\tilde \varepsilon }_{{\rm{RI}}}}{\alpha _i}{\alpha _r}} \right){\mathop{\rm Re}\nolimits} \left( {{\gamma _n}} \right) \le 0  \nonumber \vspace{-0.1cm}
\end{align}

\textbf{P-RI} is convex and is solved as detailed in Sec. IV-C.

\vspace{-0.25cm}
\subsection{Convergence and Complexity}
Problems \textbf{P-RM-b} and \textbf{P-RI} fulfill the same convergence properties as the problems analyzed in Sec. III-C. The only difference is that the convex problem solved at each iteration is a semidefinite program \cite{SIAM_SDP}, which can be tackled by using the interior-point method. By utilizing the same notation as that in Sec. III-C, the total arithmetic cost per iteration is ${\mathcal{O}}\left( {c^{1/2}}\left( {{c^3}v + {c^2}{v^2} + {v^3}} \right) \right)$ \cite[p. 247]{IPMs_Complexity}.

\vspace{-0.25cm}
\section{Numerical Results}
In this section, we focus on the problems \textbf{S-RI-a} and \textbf{P-RI}, as they encompass all the others, are the most challenging to solve, and analytical solutions are not known. The algorithms are implemented in CVX. The simulation parameters are $f = 28$ GHz, $\theta_i = 0^{\circ}$, $\theta_r = 60^{\circ}$, $R_r=R_k = 100$ m, $\eta_0 = 377$ $\Omega$, $E_0 = 1$ Watt/m$^2$, $L_x = 0.5$ m, $L_y = 4.9652 \lambda$ m, $\Delta_y = \lambda/6.0420$, $N=60$. Also, $\varepsilon_{\rm{RM}} = 2 \cdot 10^{-8}$, $\varepsilon_{\rm{RI}} = 10^{-2}$, $\varepsilon_{\rm{SP}} = 10^{-9}$, and the reradiation mask comprises the set of angles $[-2, 2]$ and $[-58, -62]$ with angular resolution 0.1$^{\circ}$. 

As for the trust region, the radii in (\ref{Eq:S-HC-b}\text{d}), (\ref{Eq:P-RM-b}\text{d}), (\ref{Eq:P-RM-b}\text{e}) are set to large values at the first iteration and are progressively reduced at each iteration by 1.2 for \textbf{S-RI-a} and by 1.1 for \textbf{P-RI}. If the objective does not decrease due to numerical inaccuracies, the following approach is used: the algorithm steps back to the solution and setup attained three iterations earlier, while the radii of the trust region are reduced by 1.2 for \textbf{S-RI-a} and by 1.1 for \textbf{P-RI}. At convergence, we attained ${\varepsilon_{\rm{TR}}} = 8.79 \cdot 10^{-13}$, ${\varepsilon _{{\rm{TR,}}{\boldsymbol{\gamma }}}} = 8.53 \cdot 10^{-6}$, ${\varepsilon _{{\rm{TR,}}{\boldsymbol{\Gamma }}}}= 1.02 \cdot 10^{-4}$. 

To ease the convergence of \textbf{P-RI},  (\ref{Eq:P-RM-b}\text{b}) and (\ref{Eq:P-RM-b}\text{c}) are rewritten as ${\left\| {\boldsymbol{\Gamma }} \right\|_{*}} - {f_{\rm{F}}}\left( {{\boldsymbol{\Gamma }},{\bar{\boldsymbol \Gamma }}} \right) \le {\varepsilon _{{\rm{rk1-b}}}}$ and $\left\| {\boldsymbol{\Gamma }} \right\| - {f_{\rm{V}}}\left( {{\boldsymbol{\gamma }},{\bar{\boldsymbol \gamma }}} \right) \le {\varepsilon _{{\rm{rk1-c}}}}$, with ${\varepsilon _{{\rm{rk1-b}}}} \ge 0$ and ${\varepsilon _{{\rm{rk1-c}}}} \ge 0$ small positive values. The impact is minor, as it only implies that the difference of the single singular value of ${\left\| {\boldsymbol{\Gamma }} \right\|}$ and ${\left\| {\boldsymbol{\gamma }} \right\|^2}$ is less than $\varepsilon _{{\rm{rk1-c}}} \ge 0$. Again, ${\varepsilon _{{\rm{rk1-b}}}}$ and ${\varepsilon _{{\rm{rk1-c}}}}$ are set to large values and are progressively reduced by 5 at each iteration, attaining ${\varepsilon _{{\rm{rk1-b}}}} = 1.25 \cdot 10^{-9}$, ${\varepsilon _{{\rm{rk1-c}}}} = 1.27 \cdot 10^{-7}$ at convergence.

As performance metric, we consider the reradiation pattern of the RIS, i.e., we plot the power flux $P_{{\theta}}\left(\boldsymbol{\gamma}\right)$ in \eqref{Power_flux_def} as a function of the angle of observation $\theta$, with $\boldsymbol{\gamma}$ obtained from the proposed optimization algorithms. Also, three benchmark schemes are considered: (i) \textbf{GO} is the geometric optics solution in \cite[Eq. (102)]{MDR_PIEEE}, with non-negative values of the real part of the surface impedance in \eqref{Surface_impedance_definition}; (ii) \textbf{GD} is the global design solution in \cite[Eq. (81)]{MDR_PIEEE}, with positive and negative values of the real part of the surface impedance in \eqref{Surface_impedance_definition}; and (iii) \textbf{GO-RI} is the GO solution in \cite[Eq. (102)]{MDR_PIEEE}, by setting the real part of the surface impedance equal to zero. The initial values of the Taylor series approximations in \textbf{S-RI-a} and \textbf{P-RI} are set to \textbf{GO-RI} for ensuring the feasibility of the initial point.

\begin{figure}[!t]
\center 
\includegraphics[width=0.91\columnwidth]{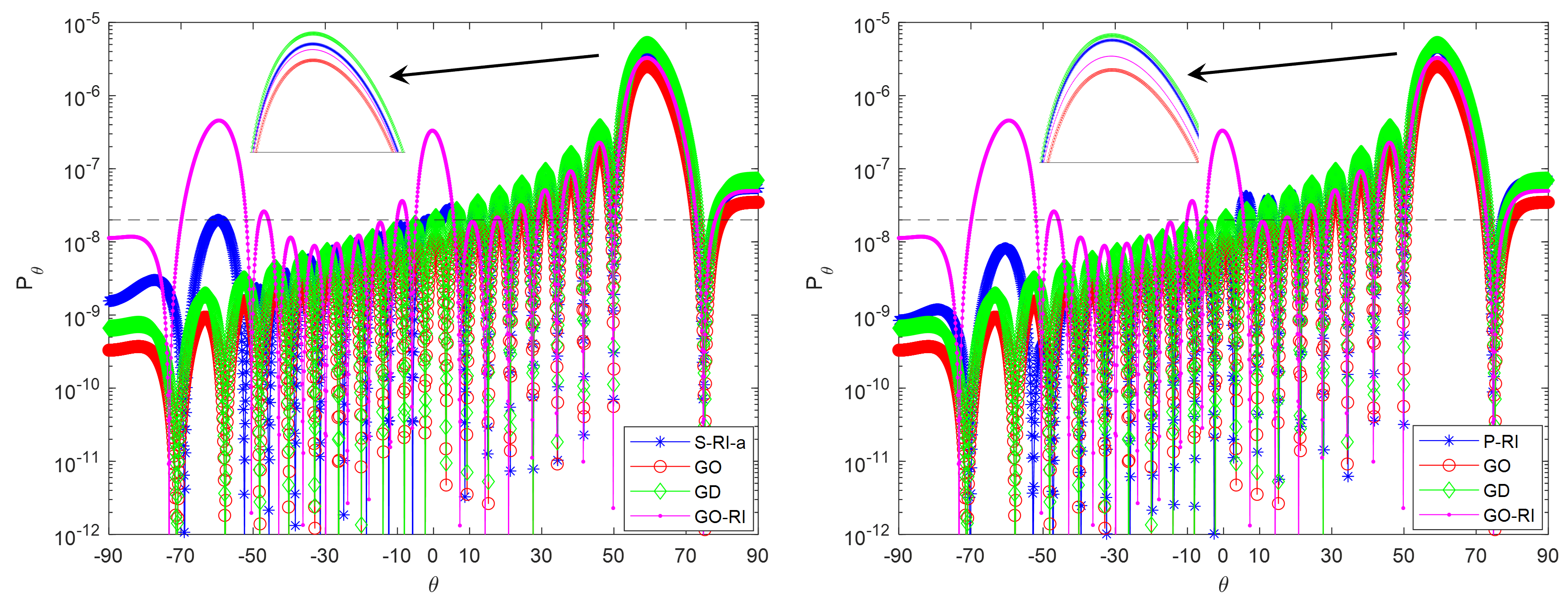} \vspace{-0.25cm}
\caption{\footnotesize Fig. 1: Reradiation pattern of the RIS: (left) \textbf{S-RI-a} and (right) \textbf{P-RI}}
\label{Fig} \vspace{-0.5cm}
\end{figure}
The results are illustrated in Fig. \ref{Fig}. According to theory \cite{MDR_PIEEE}, \textbf{GO} and \textbf{GD} provide reradiation patterns with no beams towards undesired directions of radiation, with \textbf{GD} offering the best beamforming gain at the highest implementation complexity (local amplifications due to the negative values of $\re(\bf{z})$ in \eqref{Surface_impedance_definition}). It is worth nothing that the difference of beamforming gain (at $\theta_r = 60^{\circ}$) between \textbf{GO} and \textbf{GD} is 3 dB. \textbf{GD-RI}, which is often utilized as a simple solution \cite{LoadedArrays_FMINCOM}, offers a good beamforming gain towards the direction of interest, but strong beams towards two undesired directions \cite{TAP_Floquet}. The proposed algorithms ensure a good, close to optimal, beamforming gain, no beams towards unwanted directions, and a real part of the surface impedance almost equal to zero, i.e., positive and less than $\varepsilon_{\rm{RI}}$. The problem \textbf{P-RI} usually needs higher memory requirements due to the SDP formulation (in matrix form), but it is more stable than the problem \textbf{S-RI-a} from the numerical point of view. Therefore, the proposed algorithms provide an efficient approach for optimization, ensuring close to optimal performance while fulfilling specified design constraints.

\vspace{-0.25cm}
\section{Conclusion}
We have introduced a suite of algorithms for optimizing reconfigurable anomalous reflectors based on the global design criterion. Several extensions of this work can be envisioned, including multi-beam anomalous reflecting and refracting surfaces, multiple antenna systems, near-field RIS-aided channels.

\vspace{-0.25cm}
\bibliographystyle{IEEEtran}
\bibliography{biblio}

\end{document}